\documentclass[aps,prl,twocolumn,superscriptaddress,showpacs]{revtex4}

\usepackage{graphicx}
\usepackage{subfigure}

\def\Journal#1#2#3#4{{#1} {\bf #2}, #3 (#4)}

\newcommand{\gr}{$\gamma$-ray}
\newcommand{\grs}{$\gamma$-rays}

\newcommand{\hess}{{\small H.E.S.S.}}
\newcommand{\GC}{GC}
\newcommand{\DM}{DM}
\newcommand{\MSSM}{MSSM}
\newcommand{\AMSB}{AMSB}
\newcommand{\KK}{KK}

\newcommand{\comment}[1]{}

\newcommand{\MPIK}{Max-Planck-Institut f\"ur Kernphysik, P.O. Box 103980, D 69029
Heidelberg, Germany}
\newcommand{\Yerevan}{Yerevan Physics Institute, 2 Alikhanian Brothers St., 375036 Yerevan,
Armenia}
\newcommand{\CESR}{Centre d'Etude Spatiale des Rayonnements, CNRS/UPS, 9 av. du Colonel Roche, BP
4346, F-31029 Toulouse Cedex 4, France}
\newcommand{\Hamburg}{Universit\"at Hamburg, Institut f\"ur Experimentalphysik, Luruper Chaussee
149, D 22761 Hamburg, Germany}
\newcommand{\Berlin}{Institut f\"ur Physik, Humboldt-Universit\"at zu Berlin, Newtonstr. 15,
D 12489 Berlin, Germany}
\newcommand{\LUTH}{LUTH, UMR 8102 du CNRS, Observatoire de Paris, Section de Meudon, F-92195 Meudon Cedex, France}
\newcommand{\Durham}{University of Durham, Department of Physics, South Road, Durham DH1 3LE,
U.K.}
\newcommand{\LLR}{Laboratoire Leprince-Ringuet, IN2P3/CNRS,
Ecole Polytechnique, F-91128 Palaiseau, France}
\newcommand{\APC}{APC, 11 Place Marcelin Berthelot, F-75231 Paris Cedex 05, France}
\newcommand{\Dublin}{Dublin Institute for Advanced Studies, 5 Merrion Square, Dublin 2,
Ireland}
\newcommand{\LandHD}{Landessternwarte, K\"onigstuhl, D 69117 Heidelberg, Germany}
\newcommand{\LPTA}{Laboratoire de Physique Th\'eorique et Astroparticules, IN2P3/CNRS,
Universit\'e Montpellier II, CC 70, Place Eug\`ene Bataillon, F-34095
Montpellier Cedex 5, France}
\newcommand{\LAOG}{Laboratoire d'Astrophysique de Grenoble, INSU/CNRS, Universit\'e Joseph Fourier, BP
53, F-38041 Grenoble Cedex 9, France }
\newcommand{\CEA}{DAPNIA/DSM/CEA, CE Saclay, F-91191
Gif-sur-Yvette, Cedex, France}
\newcommand{\SA}{Unit for Space Physics, North-West University, Potchefstroom 2520,
South Africa}
\newcommand{\LPNHE}{Laboratoire de Physique Nucl\'eaire et de Hautes Energies, IN2P3/CNRS, Universit\'es
Paris VI \& VII, 4 Place Jussieu, F-75252 Paris Cedex 5, France}
\newcommand{\IPNP}{Institute of Particle and Nuclear Physics, Charles University,
V Holesovickach 2, 180 00 Prague 8, Czech Republic}
\newcommand{\ITP}{Institut f\"ur Theoretische Physik, Lehrstuhl IV: Weltraum und Astrophysik,
Ruhr-Universit\"at Bochum, D 44780 Bochum, Germany}
\newcommand{\Namibia}{University of Namibia, Private Bag 13301, Windhoek, Namibia}
\newcommand{\LEA}{European Associated Laboratory for Gamma-Ray Astronomy, jointly
supported by CNRS and MPG}

\newcommand{\Purdue}{Purdue University, Department of Physics, 
525 Northwestern Avenue, West Lafayette, IN 47907-2036, USA}
\newcommand{\Tubingen}{Institut f\"ur Astronomie und Astrophysik, 
Universit\"at T\"ubingen, Sand 1, 72076 T\"ubingen, Germany}

\begin{document}


\title{H.E.S.S. observations of the Galactic Center region \\
and their possible dark matter interpretation}


\author{}
\affiliation{}


\author{F. Aharonian}
	\affiliation{\MPIK}
\author{A.G.~Akhperjanian} 
	\affiliation{\Yerevan}
\author{A.R.~Bazer-Bachi}
	\affiliation{\CESR}
\author{M.~Beilicke} \affiliation{\Hamburg}
 \author{W.~Benbow} \affiliation{\MPIK}
 \author{D.~Berge} \affiliation{\MPIK}
 \author{K.~Bernl\"ohr} \altaffiliation[Also at ]{\Berlin} \affiliation{\MPIK}
 \author{C.~Boisson} \affiliation{\LUTH}
 \author{O.~Bolz} \affiliation{\MPIK}
% ...
 \collaboration{H.E.S.S. collaboration} \homepage{http://www.mpi-hd.mpg.de/hfm/HESS/HESS.html} \noaffiliation

\date{\today}

\begin{abstract}
The detection of \grs\ from the source HESS\,J1745$-$290
in the Galactic Center (\GC) region with the \hess\ array of Cherenkov telescopes in 2004 is presented.
After subtraction of the diffuse \gr\ emission from the \GC\ ridge,
the source is compatible with a point-source with spatial extent
less than $1.2'\mbox{(stat.)}$ (95\% CL).
The measured energy spectrum above 160~GeV is compatible with
a power-law with photon index of $2.25 \pm 0.04 \mbox{(stat.)} \pm 0.10 \mbox{(syst.)}$ 
and no significant flux variation is detected.
It is finally found that the bulk of the VHE emission must have non-dark-matter origin.
\end{abstract}


\maketitle


\section{Introduction}
Recently, the CANGAROO~\cite{cangaroo}, VERITAS~\cite{veritas},
H.E.S.S.~\cite{hess} and MAGIC~\cite{magic} collaborations 
have reported the detection of very high 
energy (VHE) \grs\ in the TeV energy range from the direction of the
Galactic Center (\GC). The nature of this source is still unknown.
The main astrophysical explanations are based on particle acceleration
in the region of the Sgr~A~East supernova remnant~\cite{melia},
in the vicinity of the supermassive black hole 
Sgr~A$^*$ located at the center of our galaxy~\cite{aharonian,dermer},
or in a recently detected plerion~\cite{wang}.
Another widely discussed possibility concerns \gr\ emission
from annihilation of dark matter (\DM) particles~\cite{dmreview}.

Cosmological simulations of hierarchical structure formation~\cite{dm_simulation1,dm_simulation2}
predict that the \DM\ particles form large scale structures in the Universe, and
especially halos with a pronounced density cusp located at their center. 
Galaxies are predicted to be embedded in such \DM\ halos. 
Particle physics and cosmology experiments constrain some characteristics 
of the new particles~\cite{bertone}: the new particles should be massive ($\geq$~some GeV)
and have weak interactions with ordinary matter of the order of the electro-weak cross sections.

Extensions of the standard model of particle physics provide new particle candidates
consistent with cosmological \DM\ and are of main interest to solve both issues.
These models include supersymmetric theories 
(e.g. MSSM\footnote{MSSM: Minimal Supersymmetric Standard Model.}~\cite{ellis}
or AMSB\footnote{AMSB: Anomaly Mediated Supersymmetry Breaking.}\cite{amsb})
or Kaluza-Klein (\KK) scenarios with extra-dimensions~\cite{servant}.\\
All \DM\ particle candidates have some common properties that can be used
to detect them indirectly, since their annihilation may give rise to \grs,
but also to neutrinos and cosmic-rays. 
Their annihilation rate is proportional to the square density of \DM. 
It is thus enhanced in the dense \DM\ regions at the center of \DM\ halos.
Cuspy halos may therefore provide detectable fluxes 
of VHE \grs\ (see~\cite{bertone} and references therein).
The centers of galaxies are indeed good candidates for indirect \DM\ detection,
the closest candidate being the center of the Milky Way.
The \gr\ energy spectrum generated by \DM\ annihilation is characterized by
a continuum ranging up to the mass of the \DM\ particle,
and possibly faint \gr\ lines provided by two-body final 
states~\cite{dmreview,ellis}.

For annihilation of \DM\ particles of mass $m_{\mathrm{DM}}$ accumulated in a
spherical halo of mass density profile $\rho(r)$ and particle density profile
$\rho(r)/m_{\mathrm{DM}}$, the \gr\ flux $F(E)$ is proportional to the
line-of-sight-integrated squared particle density, multiplied by the
velocity-weighted annihilation cross section $\left<\sigma v\right>$
and the number of photons $\mbox{d}N_\gamma/\mbox{d}E$ generated per annihilation
event~\cite{darksusy}. $F(E)$ can be factored into a term $J$ depending on the halo parameters
and a term depending on the particle physics model:
$$
F(E) = F_0\,\frac{\mbox{d}N_{\gamma}}{\mbox{d}E}\,\frac{\left<\sigma v\right>}{3\cdot10^{-26}\,\mathrm{cm^3}s^{-1}}
\bigg(\frac{1\mathrm{TeV}}{m_{\mathrm{DM}}} \bigg)^2\,\bar{J}(\Delta\Omega) \Delta\Omega
$$
with $F_0 = 2.8\cdot10^{-12}\,\mathrm{cm^{-2}s^{-1}}$.
$\bar{J}(\Delta\Omega)$ denotes the ave\-ra\-ge of $J$ over the solid angle $\Delta\Omega$
corresponding to the angular resolution of the instrument, 
normalized to the local \DM\ density $0.3\,\mathrm{GeV\,cm^{-3}}$:
$$
J  = \frac{1}{8.5\,\mathrm{kpc}}\bigg(\frac{1}{0.3\,\mathrm{GeV\,cm^{-3}}} \bigg)^2
\,\int_{\ell=0}^{\infty}\mbox{d}\ell\rho^2(\ell).
$$
The shape of the measured \gr\ spectrum depends only on the particle properties,
embedded in the term $F(E)$, and especially on the \gr\ multiplicity $\mbox{d}N_\gamma/\mbox{d}E$.
The measured angular distribution of the \grs\ depends only on 
$\bar{J}(\Delta\Omega)$.
The overall \gr\ flux depends on both terms.\\
Close to the \GC, halo density profiles are predicted 
to follow a power-law $\rho_H(r) \sim r^{-\alpha}$ 
with $\alpha$ between 1~\cite{dm_simulation1} and 1.5~\cite{dm_simulation2}.
Recent N-body simulations~\cite{dm_simulation3} suggest that $\alpha$ 
could monotonically decrease to zero towards the \GC.
Values of $\alpha$ lower than $\sim1.2$ lead to an angular distribution
broader than the \hess\ angular resolution and can thus be constrained.

In this letter, we present results on VHE \grs\ from the \GC\ based
on a dataset collected in 2004 with the complete \hess\ array of 
imaging atmospheric Cherenkov telescopes (IACTs).

\section{The H.E.S.S. telescopes and the Galactic Center data set}
\hess\ is a system of four IACTs (see~\cite{Crab} and references therein) 
located in Namibia, close to the Tropic of Capricorn. The telescopes stand at the corners
of a square of 120~m side. The Cherenkov light
emitted by $\gamma$-induced air showers is imaged onto cameras of 960 photomultipliers,
covering a field of view of $5^{\circ}$ in diameter. The large mirror area of 
107~m$^2$ per telescope results in an energy threshold of 100~GeV
at Zenith~\cite{Crab}\footnote{The energy threshold is defined as the peak of the differential reconstructed 
energy distribution for a \gr\ source with a power-law energy spectrum with photon index 2.6.}.
The stereoscopic imaging of air showers 
allows the precise reconstruction of the direction and energy of the \grs.

The previously published \hess\ results on the \GC~\cite{hess}
were based on 17~h of data recorded with the first two telescopes
in 2003. Here, we report on results obtained with the full four-telescope
array, using 48.7~h (live time) of data collected between March 30th
and September 4th, 2004. The full array provides
higher detection rates than the 2003 data, 
as well as improved background rejection and angular resolution.
The bulk of the data (33.5~h) were obtained in ``wobble mode'', where the
source region is displaced by typically $\pm0.7^{\circ}$ from the optical 
axis of the system.
An additional 15.2~h dataset was obtained from the Galactic plane survey~\cite{galacticsurvey},
within  $2^{\circ}$ of Sgr~A*.

Two different techniques for calibration and image analysis were applied~\cite{Crab,model}
and give identical results.
Both methods provide a typical energy resolution of 15\%
and an angular resolution of $0.1^{\circ}$ above the analysis energy threshold.
The results described in this paper are derived using the second technique.

\section{TeV $\gamma$-rays from the direction of the Galactic Center}
\begin{figure}
\includegraphics[width=.48\textwidth]{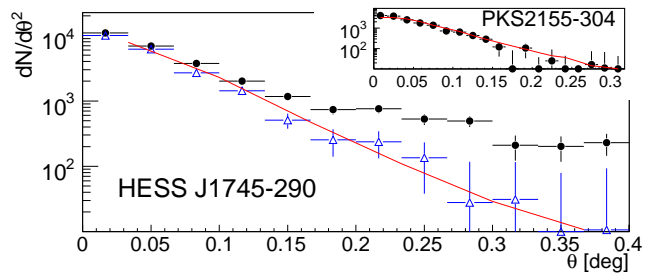}
\caption{ \label{fig:excess}
(Color online)
Background-subtracted distribution of the angle $\theta$ between the
\gr\ direction and the position of Sgr~A*.
Circles:  all detected \grs\ events.
Open triangles: central object after subtraction of the \gr\ diffuse emission model (see text).
Line: calculated PSF normalized to the number of \grs\ within $0.1^{\circ}$ after subtraction is also shown.
The distribution of events after subtraction matches the calculated PSF
while the initial distribution shows a significant tail.
The variation of the PSF related to the source energy spectrum, zenith angle and
offset position in the field of view are taken into account.
Insert: same distribution for the point-like source PKS\,2155-304~\cite{PKS2155}.
The calculated PSF (line) also matches the data.
}
\end{figure}

The data show an excess of 1863 $\gamma$-events from HESS\,J1745$-$290
within $0.1^{\circ}$ from the \GC\ (see Fig.~\ref{fig:excess}~\footnote{
From a $\mathrm{d}N/\mathrm{d}\theta$ excess distribution, the y-values have been divided 
by $2\theta$ ($\theta$ being the bin center).}). 
This excess is detected on top of a hadronic background of 1698 events, 
with a significance of 37.9~standard deviation above background, 
calculated according to~\cite{lima}.

Diffuse \gr\ emission extended along the galactic plane 
has been discovered in these data and was reported elsewhere~\cite{GCdiffuse}. 
It was shown that this emission likely originates in cosmic-ray interactions with giant molecular clouds
and is thus proportional to the density of cosmic-rays and of target material.
To study the shape and position of HESS\,J1745$-$290, the diffuse emission has been modeled 
assuming a perfect correlation with the molecular cloud density from CS data~\cite{Tsuboi}.
Cosmic-ray density was assumed to have a Gaussian dependence on distance to the GC with scale $\sigma=0.8^{\circ}$.
The resulting emission model has been smeared with the \hess\ PSF (point spread function, 
approximately Gaussian with a 68\% containment radius of $0.1^{\circ}$).
The \hess\ central source has been fitted as a superposition of the diffuse component and
either a point-like source, a Gaussian source or a \DM\ halo shape.
Likelihood fits of these different models to the \gr\
count-map within a radius of $0.5^{\circ}$ of Sgr~A$^*$
were made  with the flux normalisation of the diffuse emission model as a free
parameter.
Assuming a point source for HESS\,J1745$-$290, folded with the \hess\ PSF, 
the best fit location of the source is
($\ell=359^{\circ}56'33.3''\pm9.7''$, $b=-0^{\circ}2'40.6'' \pm 10''$) in Galactic coordinates
or ($\mathrm{\alpha=17^{h}45^{m}39.44^{s}\pm0.6^{s}}$, $\mathrm{\delta=-29d00'30.3''\pm9.7''}$) 
in equatorial coordinates (J2000.0),
within $7''\pm14''_{\mathrm{stat}}\pm28''_{\mathrm{syst}}$ 
from the putative supermassive black hole Sgr~A$^*$.
Improvements in the pointing accuracy may
allow the systematic errors to be reduced in the future. 
No remaining contribution is found in the $\gamma$-ray map after subtraction of the fitted emission,
indicating that this model is consistent with the data.
The distribution of the angle $\theta$ between the \gr\ direction and the position of Sgr~A* 
after subtraction of the fitted diffuse emission
is shown in Fig.~\ref{fig:excess} and is consistent with the \hess\ PSF.
The diffuse emission is found to contribute to 16\% of the total signal
of HESS\,J1745$-$290 within $0.1^{\circ}$.
Assuming a azimuthally symmetric Gaussian brightness distribution 
centered on the best fit position given above, folded by the \hess\ PSF,
an upper limit on the source size of 1.2' (95\% CL) was derived
(including statistical errors only).

The compatibility of the spatial extension of HESS\,J1745$-$290 with a \DM\ halo
centered on Sgr~A* and with density following $\rho(r)\propto r^{-\alpha}$
was also tested.
Different values of the logarithmic slope $\alpha$ were assumed.
The diffuse component and the \DM\ halo were both folded with the \hess\ PSF.
Leaving both normalisations free, the fit likelihood is compared to
the point-like source hypothesis discussed above in order to derive
a lower limit on the slope $\alpha$ of 1.2 (95\% CL).

\begin{figure}
\includegraphics[width=.48\textwidth]{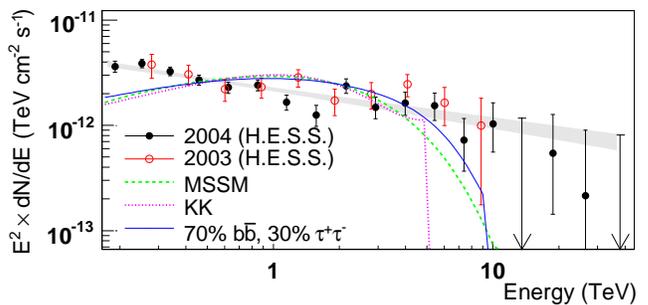}
\caption{\label{fig:edis} 
(Color online)
Spectral energy density $E^2\times \mathrm{d}N/\mathrm{d}E$ of \grs\ from 
the \GC\ source, for the 2004 data (full points) and 2003 data~\cite{hess} (open points).
Upper limits are 95\% CL.
The shaded area shows the power-law fit $\mathrm{d}N/\mathrm{d}E \sim E^{-\Gamma}$.
The dashed line illustrates typical spectra of phenomenological \MSSM\ \DM\ annihilation 
for best fit neutralino masses of 14~TeV.
The dotted line shows the distribution predicted for 
\KK\ \DM\ with a mass of 5~TeV. 
The solid line gives the spectrum of a 10~TeV \DM\ particle
annihilating into $\tau^+\tau^-$ (30\%) and $b\bar{b}$ (70\%). 
}
\end{figure}
The spectral energy distribution (SED) of \grs\ $F(E)$ of the
\GC\ source is determined using an $0.1^{\circ}$ integration radius and 
assuming a point source. 
As the flux contamination of the diffuse emission (16\%) is of the same order as flux systematic errors,
it is not subtracted in this analysis.
Moreover, as the shape of the diffuse emission spectrum is compatible with that of the
central source~\cite{GCdiffuse}, the measured spectral shape is not altered. 
The SED is shown in Fig.~\ref{fig:edis} (together with the spectrum derived from the
H.E.S.S. 2003 data).
Although a \gr\ excess is seen at energies as low as 100~GeV, the spectrum shown
is calculated only above 160~GeV to eliminate systematic errors arising from
an energy reconstruction bias close to threshold.
Over the energy range $160~\mathrm{GeV}\--30~\mathrm{TeV}$
the energy spectrum can be characterized by a power-law, 
$F(E) \sim E^{-\Gamma}$ with $\Gamma = 2.25 \pm 0.04 \mbox{(stat.)} \pm 0.10 \mbox{(syst.)}$ 
(with a fit probability of 39\%).
The 2003 and 2004 spectra are 
consistent in shape and normalization, with an integral flux 
above 1~TeV of $(1.87\pm0.10\mbox{(stat.)}\pm0.30\mbox{(syst.)})\times10^{-12}$\,cm$^{-2}$s$^{-1}$.
There is no evidence for a cut-off in the spectrum 
and lower limits at 95\% CL of 9~TeV and 7~TeV are derived 
assuming an exponential cut-off and a sharp cut-off~\footnote{$\mathrm{d}N/\mathrm{d}E=0$ above the cut-off energy.},
respectively.
The experimental spectrum has also been fitted as a sum of a free power-law and 
a monoenergetic \gr\ line
\footnote{The monoenergetic \gr\ line shape is estimated as
the \hess\ energy resolution. Its variation related to the line-energy, zenith angle, and position
of the source in the field of view, are taken into account.}  
whose energy and normalisation have been scanned.
No indications of line emission are found.

There is no significant variation in flux between 2003 and 2004;
data are consistent with a constant flux~\cite{variability}. Searches for variability
or flares on time scales down to 10~min did not show statistically
significant deviations from the mean flux. 
We note that approximately 20~min of data are required for
a $3$~standard deviation detection of the source above background.
A flare lasting for 10~min (30~min, 3~h, respectively) and with a 7-fold
(4-fold, 2-fold, respectively) increase over the quiescent flux would be detected
at the 99\%~CL.
Data were also analyzed for periodic or quasi-periodic variations 
on scales between 1~mHz and 16~$\mu$Hz, using the Lomb-Scargle method~\cite{LombScargle}.
Again, no statistically significant periodicity was found.
However, if the VHE emission is associated with Sgr~A$^*$ 
and given its typical rate of X-ray flares of 1.2 per 24~h~\cite{Chandra},
the 48.7~h of \hess\ data may simply not contain a flare event.

\section{Dark-matter interpretation}
The location of the TeV \gr\ signal and its temporal stability 
are consistent with a \DM\ annihilation signal from a halo centered Sgr~A$^*$.

In a first step, it is assumed that all \grs\ from HESS~J1745-290 are due to \DM\ annihilations.
The hypothetical \DM\ halo centered on Sgr~A* was found in the previous section to be very cuspy,
with a logarithmic slope $\alpha$ higher than 1.2.
This value is consistent with the \DM\ halo shapes
predicted by some structure formation simulations.
The energy spectrum provides another crucial test concerning a possible
\DM\ origin for the detected VHE emission. The extension of the spectrum beyond 10~TeV
requires masses of \DM\ particles which are uncomfortably large \MSSM.
The annihilation spectra of phenomenological \MSSM\ neutralinos depend
on the gaugino/higgsino mixing, but all exhibit a curved spectrum, 
which in a $E^2 \mbox{d}N/\mbox{d}E$
representation rises for $E \ll m_{\mathrm{DM}}$, plateaus at $E/m_{\mathrm{DM}} \approx 0.01\--0.1$,
and falls off approaching $m_{\mathrm{DM}}$.
\AMSB\ models lead to similar spectra.
Such a spectral shape is inconsistent
with the measured power-law as seen in Fig.~\ref{fig:edis}.
\hess\ data from 2003, with restricted energy range and lower statistics, were still marginally consistent 
with \DM\ spectra~\cite{astroph0408192}, but it appears impossible to generate a 
power-law extending over two decades from the quark and gluon fragmentation spectra 
of neutralino decays, also considering radiative effects~\cite{bergstrom}.
As an alternative scenario, mixed $\tau^+\tau^-$, $b\bar{b}$ 
final states have been proposed~\cite{Profumo},
with \DM\ masses in the $6\--30$~TeV range, generating a flatter spectrum.
Non-minimal SUSY models can be constructed which allow such decay branching ratios,
combined with neutralino masses of tens of TeV.
\KK\ \DM\ discussed in~\cite{astroph0410359} 
also give harder spectra.
PYTHIA 6.225\cite{pythia} was used to compute the contributions 
from all annihilation channels~\footnote{
Comparison of different versions of PYTHIA has shown that there are 
systematic uncertainties of the order of 10\% in the fluxes and the spectral shapes.}.
However, all the tested model spectra still deviate significantly 
from the observed power-law spectrum as shown in Fig.~\ref{fig:edis}.

On the other hand, if the bulk of the VHE emission has non-\DM\ origin,
there is still the possibility of a \DM\ signal
hidden under an astrophysical spectrum.
To search for such a contribution,
we fitted the experimental spectrum as the sum of a power-law with free 
normalization and index, and a \MSSM\ (or \KK) spectrum. Leaving the normalisation
of the \DM\ signal free, the range of $m_{\mathrm{DM}}$ is scanned. 
For the \MSSM, annihilation spectra 
$\mbox{d}N_{\gamma}/\mbox{d}E = N_0/m_{\chi}\big(E/m_{\chi}\big)^{-\Gamma}\exp\big(-c\,E/m_{\chi}\big)$
are used with three different sets of parameters, 
one approximating the average annihilation spectrum ($(N_0,\Gamma,c)=(0.081,2.31,4.88)$)
the other two ($(N_0,\Gamma,c)=(0.2,1.7,10)$ and $(0.4,1.7,3.5)$)
roughly encompassing the range of model spectra generated using Dark~Susy~\cite{darksusy}
\footnote{DarkSUSY version 4.1.}.
No significant \DM\ component is detected with this procedure,
the \DM\ component flux upper limit being of the order of 10\% of the source flux.
Assuming a NFW profile,
99\% CL upper limits on the velocity-weighted annihilation cross section 
$\left<\sigma v\right>$ are of the order of $10^{-24}\--10^{-23}\,\mathrm{cm^3s^{-1}}$, 
above the predicted values of the order of $3\times10^{-26}\,\mathrm{cm^3s^{-1}}$.
These limits can vary by plus or minus three orders of magnitude if one assumes
other \DM\ halo shapes. 
In the case of adiabatic compression of \DM\ due to the infall of baryons to the \GC,
the flux could be boosted up to a factor $1\,000$~\cite{Prada2004}.
The \hess\ data might then start to exclude some $\left<\sigma v\right>$ values.\\

In conclusion, the power-law energy spectrum
of the source HESS\,J1745-290 measured using the \hess\ telescopes
show that the observed VHE \gr\ emission is not compatible
with the most conventional \DM\ particle annihilation scenarios.
It is thus likely that the bulk of the emission is provided by astrophysical
non-\DM\ processes. However, due to high density of candidate objects
for non-thermal emission within the source region the nature of the source is not clear.



\begin{thebibliography}{99}
%
\bibitem{cangaroo} K. Tsuchiya et al. (CANGAROO Collaboration), \Journal{ApJ}{606}{L115}{2004}.
%
\bibitem{veritas} K. Kosack et al. (VERITAS Collaboration), \Journal{ApJ}{608}{L97}{2004}.
%
\bibitem{hess} F. Aharonian et al.(HESS Collaboration), \Journal{A\&A}{425}{L13}{2004}. 
%
\bibitem{magic} J. Albert et al. (MAGIC Collaboration), \Journal{ApJ}{638}{L101}{2006}.
%
\bibitem{melia} R.M. Crocker et al., \Journal{ApJ}{622}{892}{2005}. 
%
\bibitem{aharonian} F. Aharonian \& A. Neronov, \Journal{ApJ}{619}{306}{2005}. 
%
\bibitem{dermer} A. Atoyan \& C.D. Dermer, \Journal{ApJ}{617}{L123}{2004}.
%
\bibitem{wang} Q.D. Wang, \Journal{MNRAS}{367}{937}{2006}.
%
\bibitem{dmreview} L. Bergstr\"om, \Journal{Rep. Prog. Phys.}{63}{793}{2000}. 
%
%
\bibitem{dm_simulation1} J.F. Navarro, C.S. Frenk \& S.D.M. White, \Journal{ApJ}{490}{493}{1997}.
%
\bibitem{dm_simulation2} B. Moore et al., \Journal{MNRAS}{310}{1147}{1999}.
%
\bibitem{bertone} G. Bertone, D. Hooper \& J. Silk, \Journal{Phys. Rep.}{405}{279}{2005}. 
%
\bibitem{ellis} J. Ellis et al., \Journal{Eur. Phys. J.}{C24}{311}{2002}. 
%
\bibitem{amsb} S. Profumo \& P. Ullio, \Journal{JCAP}{07}{006}{2004}. 
%
\bibitem{servant} G. Servant \& T.M. Tait, \Journal{Nucl. Phys.}{B650}{391}{2003}. 
%
%
%
\bibitem{darksusy} P. Gondolo et al., \Journal{JCAP}{07}{008}{2004}.
%
\bibitem{dm_simulation3} J.F. Navarro et al., \Journal{MNRAS}{355}{794}{2004}.
%
%
%
%
\bibitem{Crab} F. Aharonian et al., \Journal{Astron. Astrophys.}{457}{899}{2006}. 
%
%
\bibitem{galacticsurvey} F. Aharonian et al., \Journal{Science}{307}{1938}{2005}. 
%
%
%
\bibitem{model} L. Rolland \& M. de Naurois, \Journal{AIP Conf. Proc.}{745}{715}{2004}.
%
\bibitem{lima} T. Li \& Y. Ma, \Journal{ApJ}{272}{317}{1983}.
%
\bibitem{GCdiffuse} F. Aharonian et al., \Journal{Nature}{439}{695}{2006}.
%
\bibitem{Tsuboi} M. Tsuboi, et al., \Journal{ApJS}{120}{1}{1999}.
%
\bibitem{PKS2155} F. Aharonian et al., \Journal{Astron. Astrophys.}{442}{895}{2005}.
%
\bibitem{variability} L. Rolland et al., Proc. 29th ICRC, Pune (2005). 
%
\bibitem{LombScargle} J.D. Scargle, \Journal{ApJ}{343}{874}{1989}.
%
\bibitem{Chandra} M.P. Muno et al., \Journal{ApJ}{589}{225}{2003}.  
%
\bibitem{astroph0408192} D. Horns, \Journal{Phys. Lett.}{B607}{225}{2005}.    
%
%
\bibitem{bergstrom} L. Bergstr\"om et al., \Journal{Phys. Rev. Lett.}{95}{241301}{2005}. 
%
\bibitem{Profumo} S. Profumo, \Journal{Phys. Rev.}{D72}{10352}{2005}.  
%
\bibitem{astroph0410359} L. Bergstr\" om et al., \Journal{Phys. Rev. Lett.}{94}{131301}{2005}. 
%
%
\bibitem{pythia} T. Sj\"ostrand et al., \Journal{Computer Phys. Commun.}{135}{238}{2001}.
%
%
\bibitem{Prada2004} F. Prada et al., \Journal{Phys. Rev. Lett.}{93}{241301}{2004}.
\end{thebibliography}

\end{document}